\documentclass[prd,showpacs,amsmath,amssymb,superscriptaddress,preprintnumbers,nofootinbib,showkeys]{revtex4}

\begin{document}

\newcommand{\nablab}{{\mathop {\rule{0pt}{0pt}{\nabla}}\limits^{\bot}}\rule{0pt}{0pt}}

\title{Dark Energy and Dark Matter Interaction:  \\ Kernels of Volterra Type and Coincidence Problem}

\author{Alexander B. Balakin}
\email{Alexander.Balakin@kpfu.ru}
\author{Alexei S. Ilin}
\email{alexeyilinjukeu@gmail.com}
\affiliation{Department of General Relativity and
Gravitation, Institute of Physics,Kazan Federal University, Kremlevskaya street 18, Kazan, 420008,
Russia}
\date{\today}

\begin{abstract}
We study a new exactly solvable model of coupling of the Dark Energy and Dark Matter, in the framework of which the kernel of non-gravitational interaction is presented by the integral Volterra-type operator well-known in the classical theory of fading memory. Exact solutions of this isotropic homogeneous cosmological model were classified with respect to the sign of the discriminant of the cubic characteristic polynomial associated with the key equation of the model. Energy-density scalars of the Dark Energy and Dark Matter, the Hubble function and acceleration parameter are presented explicitly; the scale factor is found in quadratures. Asymptotic analysis of the exact solutions has shown that the Big Rip, Little Rip, Pseudo Rip regimes can be realized with  the specific choice of guiding parameters of the model. We show that the Coincidence problem can be solved if we consider the memory effect associated with the interactions in the Dark Sector of the Universe.
\end{abstract}
\maketitle

\section{Introduction}

Dark Matter (DM) and Dark Energy (DE) play the key roles in all modern cosmological scenaria (see, e.g., \cite{DM1} - \cite{DMDE4}, and references therein for the history of problem, for main ideas and mathematical details). The Dark Matter and Dark Energy interact by the gravitational field, thus creating the space-time background for various astrophysical and cosmological events. In addition, according to the general view, the direct (non-gravitational) DM/DE coupling exists. One of the motivation of this idea is connected with the so-called Coincidence Problem \cite{CO1,CO2,CO3}), which is based on the fact that the ratio between DE and DM energy densities is nowadays of the order $\frac{73}{23}$, while at the Planck time this ratio was of the order $10^{-95}$, if one uses for calculations the energy density, associated with the cosmological constant (see, e.g., the review \cite{Pavon2} for details of estimations). Clearly, the non-gravitational interactions between the DE and DM, or for short, interactions in the Dark Sector of the Universe, could start up the self-regulation procedure thus eliminating the initial disbalance. There is a number of models, which describe the DE/DM coupling (see, e.g., \cite{i1,Z,Pavon1,Pavon2}). The most known models are phenomenological; they operate with the so-called kernel of interaction, the function $Q(t)$, which is linear in the energy densities of the Dark Energy and Dark Matter with coefficients proportional to the Hubble function \cite{Pavon2}. In the series of works \cite{Arc1,Arc2,Arc3,Arc4} the DE/DM interaction is modeled on the base of relativistic kinetic theory with an assumption that DE acts on the DM particles by the gradient force of the Archimedean type. In \cite{BD2014,Obzor} the DE/DM interactions are considered in terms of extended electrodynamics of continua. In \cite{QQ} the kernel $Q(t)$ was reconstructed for the case, when the cosmological expansion is described by the hybrid scale factor, composed using both: power-law and exponential functions.

In this work we present the function $Q(t)$, the kernel of DE/DM interaction, in the integral form, using the analogy with classical theory of fading memory. The appropriate mathematical formalism is based on the theory of linear Volterra operators \cite{Volterra}; the corresponding integrand contains the difference of the DE and DM energy densities. The kernel of interaction vanishes if the DE and DM energy densities coincide; when these quantities do not coincide, the kernel of interaction acts as the source in the balance equations providing the procedure of self-regulation. However, in contrast to the known local phenomenological representations of the interaction kernels, the value of the source-function $Q(t)$ in the model, which includes the Volterra integrals, is predetermined by whole prehistory of the Universe evolution. As the result of modeling, we see that the ratio between the DE and DM energy densities tends asymptotically to some theoretically predicted value, which can be verified using the cosmological observations.

The paper is organized as follows. In Section 2 we recall the main elements of the phenomenological approach to the Universe evolution filled by interacting Dark Energy and Dark Matter. In Section 3 we formulate the model with kernel of the Volterra type, derive the integro-differential equations describing the Universe evolution, and obtain the so-called key equation, which is the differential equation of the Euler type of the third order in ordinary derivatives for the DE energy density. In Section 4 we classify the exact solutions with respect to the sign of the discriminant of the characteristic polynomial. In Section 5 we consider three examples of explicit analysis of the Universe evolution in the proposed model, and distinguish two exact solutions indicated as bounce and super-inflation, respectively. Section 6 contains discussion and conclusions.

\section{Phenomenological approach to the problem of interactions in the Dark Sector of the Universe}

First of all we would like to recall how do the phenomenological elements appear in the theory of interactions in the Dark Sector of the Universe. We consider the well-known two-fluid model, which describes the so-called Dark Fluid joining the
Dark Energy and Dark Matter; in this model the baryonic matter remains out of consideration.

\subsection{Two-fluid model in the Einstein theory of gravity}

The master equations for the gravity field
\begin{equation}
R^{ik}-\frac12 g^{ik} R - \Lambda g^{ik} = \kappa	\left[T^{ik}_{(\rm DE)} + T^{ik}_{(\rm DM)} \right]
\label{1}
\end{equation}
are considered to be derived from the Hilbert-Einstein action functional. Here $R^{ik}$ is the Ricci tensor; $R$ is the Ricci scalar; $\Lambda$ is the cosmological constant; $T^{ik}_{(\rm DE)}$ and $T^{ik}_{(\rm DM)}$ are the stress-energy tensors of the Dark Energy and Dark Matter, respectively. These tensors can be algebraically decomposed using the Landau-Lifshitz scheme of definition of the fluid macroscopic velocity:
\begin{equation}
T^{ik}_{(\rm DE)} = W U^i U^k + {\cal P}^{ik} \,,
\quad
T^{ik}_{(\rm DM)} = E V^i V^k + \Pi^{ik} \,.
\label{5}
\end{equation}
Here $U^i$ and $V^i$ are the timelike velocity four-vectors, the eigen-vectors of the corresponding stress-energy tensors:  \begin{equation}
U_i T^{ik}_{(\rm DE)} = W U^k \,, \quad V_k T^{ik}_{(\rm DM)} = E V^i \,.
\label{7}
\end{equation}
The corresponding eigen-values, the scalars $W$ and $E$ are the energy density scalars of DE and DM, respectively.
The quantities ${\cal P}^{ik}$ and $\Pi^{ik}$ are the pressure tensors of the DE and DM; they are orthogonal to the velocity four-vectors:
\begin{equation}
U_i {\cal P}^{ik} = 0 \,, \quad  V_k \Pi^{ik}=0 \,.
\label{2}
\end{equation}
The Bianchi identity provides the sum of the DE and DM stress-energy tensors to be divergence free:
\begin{equation}
\nabla_k\left[T^{ik}_{(\rm DE)} + T^{ik}_{(\rm DM)} \right]=0	 \,.
\label{8}
\end{equation}
This means that there exists a vector field $Q^i$, which possesses the property
\begin{equation}
\nabla_k T^{ik}_{(\rm DE)} = Q^i = - \nabla_k T^{ik}_{(\rm DM)}  \,.
\label{11}
\end{equation}
Till now we did not use the phenomenological assumptions; only the next step, namely the modeling of the vector field $Q^i$ is the essence of the phenomenological approach, which describes the DE/DM interactions.

\subsection{Description of the DE/DM coupling in the framework of an isotropic homogeneous cosmological model}

When one deals with the spatially isotropic homogeneous cosmological model the key elements of the theory of DE and DM coupling can be simplified essentially. First of all, one uses the metric
\begin{equation}
	ds^2=dt^2-a^2(t)\left[dx^2+dy^2+dz^2 \right] \,,
\label{13}
\end{equation}
with the scale factor $a(t)$ depending on the cosmological time; one assumes that the energy-density scalars also depend on time only, $W(t)$, $E(t)$. Second, the eigen four-vectors $U^i$ and $V^i$ coincide and are of the form $U^i = V^i = \delta^i_0$. Third, the pressure tensors happen to be reduced to the Pascal-type scalars $P(t)$ and $\Pi(t)$:	
\begin{equation}
{\cal P}^{ik} = - P \Delta^{ik} \,, \quad \Pi^{ik} = - \Pi \Delta^{ik} \,, \quad \Delta^{ik} = g^{ik}-U^iU^k  \,.
\label{14}
\end{equation}
The four-vector $Q^i$ now is presented by one scalar function $Q(t)$, since $Q^i=Q U^i$ in the spatially isotropic model. The function $Q(t)$ is called in the review \cite{Pavon2} by the term {\em kernel} of interaction. The master equations of the model can be now reduced to the following three ones:
\begin{equation}
	3H^2 - \Lambda = \kappa \left[W(t)+E(t) \right] \,,
\label{15}
\end{equation}
\begin{equation}
	\dot{W}+3H(W+P)=Q \,,
	\label{16}
\end{equation}
	\begin{equation}
	\dot{E}+3H(E+\Pi)=-Q \,,
	\label{17}
\end{equation}
where $H(t) \equiv \frac{\dot{a}}{a}$ is the Hubble function, and the dot denotes the derivative with respect to time.
The equation (\ref{15}) is taken from the Einstein equations; the sum of  (\ref{16}) and (\ref{17}) gives the total energy conservation law. Also, we use the standard linear equations of state
\begin{equation}
P=(\Gamma-1) W \,, \quad \Pi=(\gamma-1) E \,,
\label{31}
\end{equation}
which allow us to focus on the analysis of the set of three equations for three unknown functions $W$, $E$ and $H$. The history of modeling of the function $Q(t)$ is well documented in the review \cite{Pavon2}; we focus below on a new (rheological-type) model.

\section{Rheological-type  model of the DE/DM coupling}

\subsection{Reconstruction of the kernel $Q(t)$}

In order to reconstruct phenomenologically the interaction kernel $Q(t)$ we use the ansatz based on the following
three assumptions.

\noindent
{\it (i)} The function $Q(t)$ is presented by the integral operator of the Volterra type:
\begin{equation}
Q(t) = \int_{t_{0}}^{t}d\xi K(t,\xi)[E(\xi)-W(\xi)]	\,.
\label{21}
\end{equation}

\noindent
{\it (ii)} The Volterra integral contains the difference of the energy density scalars $E(\xi)$ and $W(\xi)$.

\noindent
{\it (iii)} The kernel of the Volterra integral $K(t,\xi)$ has a specific multiplicative form
\begin{equation}
K(t,\xi) = K_{0} H(t) H(\xi)\left[\frac{a(\xi)}{a(t)}\right]^{\nu}\,.	
\label{22}
\end{equation}
Motivation of our choice is the following.

1) In the context of rheological approach we assume that the state of a fluid system at the present time moment $t$ is predetermined by whole prehistory of its evolution from the starting moment $t_0$ till to the moment $t$. More than century ago it was shown, that the mathematical formalism appropriate for description of this idea can be based on the theory of linear integral Volterra operators, which have found numerous applications to the theory of media with memory. We also use this fruitful idea.

2) Our ansatz is that the interaction between two constituents of the Dark Fluid vanishes, if the DE energy density coincides identically  with the DM energy density, $W \equiv E$. When $W \neq E$ the integral mechanism of self-regulation inside the Dark Fluid switches on. For instance,  during the cosmological epochs with DE domination, i.e., when $W>E$, the corresponding contribution into the interaction term $Q$ is negative, the rates $\dot{W}$ and $\dot{E}$ obtain negative and positive contributions, respectively (see (\ref{16}) and (\ref{17})); when $W<E$, the inverse process starts thus regulating the ratio between DE and DM energy densities.

3) For classical models of fading memory the kernel of the Volterra operator is known to be of exponential form
$K(t,\xi)={\cal K} \exp{\frac{(\xi-t)}{T_0}}$, where the parameter $T_0$ describes the typical time of memory fading, and the quantity ${\cal K}$ has the dimensionality $[{\rm time}]^{-2}$. When we work with the de Sitter scale factor $a(t)=a(t_0)\exp{H_0 t}$, we can rewrite the kernel of the Volterra operator as follows:
\begin{equation}
K(t,\xi) = {\cal K} \exp{\left[\frac{H_0(\xi-t)}{H_0 T_0}\right]} = K_{0} H^2_{0}\left[\frac{a(\xi)}{a(t)}\right]^{\frac{1}{H_0 T_0}}\,,	
\label{23}
\end{equation}
where the parameter $K_0$ is dimensionless.
This idea inspired us to formulate the ansatz, that not only for the de Sitter spacetime, but for Friedmann - type spacetimes also, we can use the kernel (\ref{22}) with two additional model parameters, $\nu$ and $K_0$.

\subsection{Key equation of the model}

In order to analyze the set of coupled equations (\ref{15})-(\ref{31}), (\ref{21}) with (\ref{22}), let us derive the so-called key equation, which contains only one unknown function, $W$. In our model with the ansatz (\ref{22}) the unknown functions $W$ and $E$ depend on time through the scale factor, i.e., $W=W(a(t))$, $E=E(a(t))$. Following the standard approach (see, e.g., the review \cite{Pavon2}), we  introduce new dimensionless variable $x$ instead of $t$ using the definitions
\begin{equation}
x \equiv \frac{a(t)}{a(t_0)} \,, \quad \frac{d}{dt} = x H(x)\frac{d}{dx} \,. 	
\label{25}
\end{equation}
When the function $H(x)$ is found, the scale factor as the function of cosmological time can be found from the following quadrature:
\begin{equation}
t-t_0 = \int_1^{\frac{a(t)}{a(t_0)}}\frac{dx}{x H(x)}\,.	
\label{250}
\end{equation}
In these terms three basic master equations take the form
\begin{equation}
3H^2(x)- \Lambda = \kappa \left[W(x)+E(x) \right]\,,	
\label{26}
\end{equation}
\begin{equation}
x \frac{dW}{dx}+3 \Gamma W = K_{0} x^{-\nu}\int_{1}^{x}dy y^{\nu-1} [E(y)-W(y)] \,,	
\label{27}
\end{equation}
\begin{equation}
x \frac{dE}{dx}+3\gamma E = K_{0} x^{-\nu}\int_{1}^{x}dy y^{\nu-1} [W(y)-E(y)] \,.	
\label{28}
\end{equation}
Also, we have the consequence of two last equations:
\begin{equation}
x \frac{d}{dx}(W+E)+3 \left(\Gamma W + \gamma E\right) = 0 \,.	
\label{281}
\end{equation}
The equation (\ref{26}) is decoupled from this set; it can be used to find the Hubble function, when $W(x)$ and $E(x)$ are obtained. Two last integro-differential equations can be reduced to the differential ones:
\begin{equation}
x^2 W^{\prime \prime} + x W^{\prime}(\nu + 1 + 3\Gamma) + 3 \nu \Gamma W =K_{0}(E-W) \,,
\label{41}
\end{equation}
\begin{equation}
x^2 E^{\prime \prime} + x E^{\prime}(\nu + 1 + 3\gamma) + 3 \nu \gamma E =K_{0}(W-E) \,.
\label{42}
\end{equation}
Here and below the prime denotes the derivative with respect to dimensionless variable $x$.	
The next step is the following: we extract $E(x)$ from (\ref{41})
\begin{equation}
E(x) = \frac{1}{K_{0}}\left[x^2 W^{\prime \prime} + x W^{\prime}(\nu + 1 + 3\Gamma) + W(K_0+3 \nu \Gamma) \right] \,,
\label{43}
\end{equation}
and put it into (\ref{281}), thus obtaining	the Euler equation of the third order
\begin{equation}
x^3 W^{\prime \prime \prime} + (A+3) x^2 W^{\prime \prime} + (B+1) x W^{\prime} + D W =0 \,,
\label{51}
\end{equation}
where the auxiliary parameters are the following:	
\begin{equation}
A = \nu + 3(\Gamma + \gamma) \,, \quad  B = A + 2K_0 + 3\nu (\Gamma + \gamma) + 9 \Gamma \gamma \,, \quad  D = 3\left[K_0(\Gamma + \gamma) + 3\nu \Gamma \gamma \right] \,.
\label{52}
\end{equation}
We indicate the equation (\ref{51}) as the key equation, since when $W(x)$ is found, we obtain $E(x)$ immediately from (\ref{43}), and then $H(x)$ from (\ref{26}).

\section{Classification of solutions}

\subsection{The scheme of classification}

The characteristic equation for the Euler equation (\ref{51}) is the cubic one:
\begin{equation}
\sigma^3+\sigma^2 A+\sigma (B-A)+D=0 \,.
\label{61}
\end{equation}
As usual, we reduce the cubic equation to the canonic form
\begin{equation}
\sigma = z - \frac{A}{3} \ \ \rightarrow \  z^3+pz +q=0 \,,
\label{62}
\end{equation}
using the following definitions of the canonic parameters $p$ and $q$:
\begin{equation}
p = B-A-\frac13 A^2 \,, \quad q = \frac{2}{27} A^3 + \frac13 A(A-B) + D \,.
\label{63}
\end{equation}
The discriminant of the cubic equation (\ref{61}) with $p$ and $q$ given by (\ref{63}) is of the form
\begin{equation}
\Delta = \frac{p^3}{27} + \frac{q^2}{4}  \,.
\label{71}
\end{equation}
When $\Delta< 0$, the roots of the equation (\ref{61}) are real and do not coincide, $\sigma_1 \neq \sigma_2 \neq \sigma_3$.
\noindent
When $\Delta =0$, the roots are real, but at least two of them coincide, $\sigma_1 \neq \sigma_2 = \sigma_3$ or
$\sigma_1 = \sigma_2 = \sigma_3$.
\noindent
When $\Delta >0$, there is one real root, and  a pair of complex conjugated, $\sigma_1$, $\sigma_{2,3} = \alpha \pm i \beta$.  Let us study all these cases in detail.

\subsection{Solutions corresponding to the negative discriminant, $\Delta  < 0$}

\subsubsection{The structure of the exact solution}

It is the case, when the parameter $p$ is negative, $p<0$, and $\left|\frac{q}{2}\right| \left(\frac{3}{|p|}\right)^{\frac32}  <1$, or in more detail
\begin{equation}
\left|\frac{1}{27} A^3 + \frac16 A(A-B) + \frac12 D \right| < \left|\frac13(B-A)-\frac19 A^2 \right|^{\frac32}\,.
\label{001}
\end{equation}
All three roots $\sigma_1$, $\sigma_2$, $\sigma_3$ are real and they do not coincide:
\begin{equation}
\sigma_1 = -\frac{A}{3} + 2\sqrt{\frac{|p|}{3}} \cos{\frac{\varphi}{3}}  \,, \quad
\sigma_{2,3}= -\frac{A}{3} + 2\sqrt{\frac{|p|}{3}} \cos{\left(\frac{\varphi}{3} \pm \frac{2\pi}{3}\right)} \,,
\label{01}
\end{equation}
where the auxiliary angle $0 \leq \varphi \leq \pi$ is defined as follows:
\begin{equation}
\cos{\varphi} = -\frac{q}{2} \left( \frac{3}{|p|}\right)^{\frac32} \,.
\label{02}
\end{equation}
In this case the key equation for the DE energy density scalar $W(x)$ gives power-law solution:
\begin{equation}
W(x) = C_1 x^{\sigma_1} + C_2 x^{\sigma_2} + C_3 x^{\sigma_3} \,.
\label{72}
\end{equation}
Using the relationship (\ref{43}) we obtain immediately the DM energy density scalar $E(x)$
$$
E(x)  = \frac{1}{K_0} \left\{  C_1 x^{\sigma_1}\left[\sigma^2_1 + \sigma_1 (\nu + 3\Gamma) + (K_0+3 \nu \Gamma)\right]   +
\right.
$$
\begin{equation}
\left.
 + C_2 x^{\sigma_2}\left[\sigma^2_2 + \sigma_2 (\nu + 3\Gamma) + (K_0+3 \nu \Gamma)\right]
+ C_3 x^{\sigma_3}\left[\sigma^2_3 + \sigma_3 (\nu + 3\Gamma) + (K_0+3 \nu \Gamma)\right] \right\} \,.
\label{73}
\end{equation}
The constants of integration $C_1$, $C_2$, $C_3$ can be expressed in terms of presented functions at $t=t_0$, or equivalently, at $x=1$; they are the solutions of the system:
\begin{equation}
C_1 + C_2 + C_3 = W(1) \,,
\label{721}
\end{equation}
\begin{equation}
C_1 \sigma_1 + C_2 \sigma_2 + C_3 \sigma_3 = - 3 \Gamma W(1) \,,
\label{729}
\end{equation}
$$
 C_1 \left[\sigma^2_1 + \sigma_1 (\nu + 3\Gamma) + (K_0+3 \nu \Gamma)\right]   +
 C_2 \left[\sigma^2_2 + \sigma_2 (\nu + 3\Gamma) + (K_0+3 \nu \Gamma)\right] +
 $$
\begin{equation}
+ C_3 \left[\sigma^2_3 + \sigma_3 (\nu + 3\Gamma) + (K_0+3 \nu \Gamma)\right]] = K_0 E(1)  \,.
\label{731}
\end{equation}
Clearly, the first and third equations are the direct consequences of (\ref{72}) and (\ref{73}), respectively; as for the second relationship, we obtain it from (\ref{27}), when $x=1$. The Cramer determinant for this system
\begin{equation}
{\cal D} = (\sigma_1-\sigma_2)(\sigma_2-\sigma_3)(\sigma_3-\sigma_1)  \neq 0
\label{81}
\end{equation}
is not equal to zero, thus the system has the unique solution:
\begin{equation}
C_1 = \frac{1}{(\sigma_1-\sigma_2)(\sigma_1-\sigma_3)}\left\{
W(1)\left[\sigma_2 \sigma_3 + 3\Gamma (\sigma_2 + \sigma_3) + 9\Gamma^2 -K_0 \right] + K_0 E(1)
\right\} \,,
\label{82}
\end{equation}
\begin{equation}
C_2 = \frac{1}{(\sigma_2-\sigma_1)(\sigma_2-\sigma_3)}\left\{
W(1)\left[\sigma_1 \sigma_3 + 3\Gamma (\sigma_1 + \sigma_3) + 9\Gamma^2 -K_0 \right] + K_0 E(1)
\right\} \,,
\label{83}
\end{equation}
\begin{equation}
C_3 = \frac{1}{(\sigma_3-\sigma_1)(\sigma_3-\sigma_2)}\left\{
W(1)\left[\sigma_1 \sigma_2 + 3\Gamma (\sigma_1 + \sigma_2) + 9\Gamma^2 -K_0 \right] + K_0 E(1)
\right\} \,.
\label{84}
\end{equation}
Then, using the Einstein equation (\ref{26}) we find the square of the Hubble function:
$$
 H^2(x) = \frac{\Lambda}{3} + \frac{\kappa}{3K_0} \left\{C_1 x^{\sigma_1}\left[2K_0+\sigma^2_1 + \sigma_1 (\nu + 3\Gamma) + 3 \nu \Gamma \right] +
\right.
$$
\begin{equation}
\left.
+C_2 x^{\sigma_2}\left[2K_0+\sigma^2_2 + \sigma_2 (\nu + 3\Gamma) + 3 \nu \Gamma \right] +
C_3 x^{\sigma_3}\left[2K_0+\sigma^2_3 + \sigma_3 (\nu + 3\Gamma) + 3 \nu \Gamma \right]\right\} \,.
\label{75}
\end{equation}
The scale factor $a(t)$ can be now obtained from the integral
\begin{equation}
\sqrt{\frac{\kappa}{3K_0}}(t-t_0) = \int_1^{\frac{a(t)}{a(t_0)}}\frac{dx}{x \sqrt{\frac{K_0 \Lambda}{\kappa}+\sum_{j=1}^3 C_j x^{\sigma_j}\left[2K_0+\sigma^2_j + \sigma_j (\nu + 3\Gamma) + 3 \nu \Gamma \right]}}\,.	
\label{025}
\end{equation}
Generally, this integral can not be expressed in elementary functions; results of asymptotic analysis are discussed below.

\subsubsection{Two auxiliary characteristics of the model and a scheme of estimation of the kernel parameters}

\noindent
{\it (1) The acceleration parameter $q$}

\noindent
The formula (\ref{75}) allows us to calculate immediately the acceleration parameter:
\begin{equation}
-q(x) = 1+ \frac{x}{2H^2(x)} \frac{d H^2}{dx} = \frac{\frac{K_0 \Lambda}{\kappa} + \sum_{j=1}^3 C_j x^{\sigma_j} \left(1+\frac{\sigma_j}{2} \right)\left[2K_0+\sigma^2_j + \sigma_j (\nu + 3\Gamma) + 3 \nu \Gamma \right]}{\frac{K_0 \Lambda}{\kappa} + \sum_{j=1}^3 C_j x^{\sigma_j}\left[2K_0+\sigma^2_j + \sigma_j (\nu + 3\Gamma) + 3 \nu \Gamma \right]} \,.
\label{q1}
\end{equation}

\noindent
{\it (2) The DM/DE energy density ratio $\omega$}

\noindent
For many purposes it is important to have the ratio $\omega(x) = \frac{E(x)}{W(x)}$. Direct calculation gives
\begin{equation}
\omega(x) = \frac{E(x)}{W(x)} =  \frac{\sum_{j=1}^3 C_j x^{\sigma_j} \left[\sigma^2_j + \sigma_j (\nu + 3\Gamma) + (K_0+3 \nu \Gamma) \right]}{K_0 \sum_{j=1}^3 C_j x^{\sigma_j}} \,.
\label{q12}
\end{equation}
Let us assume that the present moment of the cosmological time is $t=T$, and the corresponding value of the dimensionless scale factor is $X=\frac{a(T)}{a(t_0)}$. Also, we use the following estimations for the present time parameters:
\begin{equation}
\omega(X) = \frac{E(X)}{W(X)} \simeq \frac{23}{73} \,, \quad q(X) = - 0.55 \,.
\label{est1}
\end{equation}
Thus, we have two relationships, which link the kernel parameters $K_0$ and $\nu$ with $X$ and other coupling constants:
\begin{equation}
-0.9 = \frac{\sum_{j=1}^3 C_j X^{\sigma_j} \sigma_j \left[2K_0+\sigma^2_j + \sigma_j (\nu + 3\Gamma) + 3 \nu \Gamma \right]}{\frac{K_0 \Lambda}{\kappa} + \sum_{j=1}^3 C_j X^{\sigma_j}\left[2K_0+\sigma^2_j + \sigma_j (\nu + 3\Gamma) + 3 \nu \Gamma \right]} \,,
\label{q1X}
\end{equation}
\begin{equation}
   \frac{23}{73}=  \frac{\sum_{j=1}^3 C_j X^{\sigma_j} \left[\sigma^2_j + \sigma_j (\nu + 3\Gamma) + (K_0+3 \nu \Gamma) \right]}{K_0 \sum_{j=1}^3 C_j X^{\sigma_j}} \,.
\label{q12X}
\end{equation}
We hope to realize the whole scheme of fitting of the model parameters in a special paper.

\subsubsection{Admissible asymptotic regimes, and constraints on the model parameters}

There are three regimes of asymptotic behavior of the presented solutions.

\vspace{3mm}
\noindent
{\it (i)}
If the maximal real root, say $\sigma_1$, is positive and the set of initial data is general, we see that $W \to \infty$, $E \to \infty$ and $H \to \infty$, when $x\to \infty$. The integral in (\ref{250}) converges at $a(t) \to \infty$, and the scale factor $a(t)$ follows the law $a(t)=a_{*} (t_{*}-t)^{-\frac{2}{\sigma_1}}$, and reaches infinity at $t=t_{*}$.
We deal in this case with the so-called Big Rip asymptotic regime, and the Universe follows the catastrophic scenario \cite{CO2,O1}.
In particular,
when $\sigma_1>0$ and $\sigma_2<0$, $\sigma_3<0$, according to the Vi\`ete theorem, we can definitely say only that $\sigma_1 \sigma_2 \sigma_3 = -D>0$, i.e.,
$K_0 (\Gamma+\gamma)+ 3\nu \Gamma \gamma < 0$.
The asymptotic value of the acceleration parameter is equal to $-q(\infty) = 1+ \frac{\sigma_1}{2}$.
The final ratio between the DM and DE energy densities
\begin{equation}
\omega(\infty) = \frac{\sigma^2_1 + \sigma_1 (\nu + 3\Gamma) + K_0 +3 \nu \Gamma}{K_0}
\label{q125}
\end{equation}
does not depend on the initial parameters $W(1)$ and/or $E(1)$.

\vspace{3mm}
\noindent
{\it (ii)}
If the maximal real root, say $\sigma_1$, is equal to zero, we see that $D=0$, and thus
\begin{equation}
K_0 (\Gamma+\gamma)+ 3\nu \Gamma \gamma = 0 \,.
\label{009}
\end{equation}
In this case the Hubble function tends asymptotically to constant $H_{\infty}$, given by
\begin{equation}
H_{\infty} = \sqrt{\frac{\Lambda}{3}+\frac{\kappa \left(2K_0 + 3 \nu \Gamma \right)}{3K_0 \sigma_2 \sigma_3}\left\{
W(1)\left[\sigma_2 \sigma_3 + 3\Gamma (\sigma_2 + \sigma_3) + 9\Gamma^2 -K_0 \right] + K_0 E(1)
\right\} }\,,
\label{006}
\end{equation}
thus providing the scale factor to be of the exponential form $a(t) \to a_{\infty} e^{H_{\infty} t}$; we deal in this case with the Pseudo Rip, or in other words, the late-time Universe of the quasi-de Sitter type. Clearly, the asymptotic value of the function $-q(x)$, given by (\ref{q1}), is $-q(\infty)=1$. As for the asymptotic value of the quantity $\omega(x)$ (see (\ref{q12})), it
is now equal to $\omega(\infty)= -\frac{\Gamma}{\gamma}$. Since $\omega$ is the non-negatively defined quantity, this situation is possible only if the ratio $\frac{\Gamma}{\gamma}$ is non-positive. Thus, the evolution of the ratio $\frac{E(x)}{W(x)}$ starts from the value $\frac{E(1)}{W(1)}$ and finishes with $\left|\frac{\Gamma}{\gamma}\right|$. One can add that, when $\sigma_1=0$ and $\sigma_2<0$, $\sigma_3<0$, we obtain two supplementary inequalities:
\begin{equation}
A = - (\sigma_2 + \sigma_3) >0   \ \rightarrow \nu + 3 (\Gamma+\gamma) > 0 \,,
\label{010}
\end{equation}
\begin{equation}
B-A = \sigma_2 \sigma_3 >0   \ \rightarrow 2K_0  + 3 \nu (\Gamma+\gamma) + 9 \Gamma \gamma > 0 \,.
\label{011}
\end{equation}
These requirements restrict the choice of model parameters.

\vspace{3mm}
\noindent
{\it (iii)}
If all the roots are negative, we see that $H \to H_0 \equiv \sqrt{\frac{\Lambda}{3}}$, when $x\to \infty$, thus we obtain the classical de Sitter asymptote with $-q(\infty)=1$. When $\Lambda=0$, all the roots are negative, and, say, $\sigma_1$ is the maximal among them, we see that $W \to 0$, $E \to 0$ at $x\to \infty$. The scale factor behaves asymptotically as
the power-law function $a(t) \propto t^{\frac{2}{|\sigma_1|}}$; the acceleration parameter $-q(\infty) = 1-\frac{|\sigma_1|}{2}$ is positive, when $|\sigma_1|<2$.
In particular,
when $\sigma_1<0$, $\sigma_2<0$, $\sigma_3<0$, we see that, first, $\sigma_1 \sigma_2 \sigma_3 = -D<0$, i.e.,
$K_0 (\Gamma+\gamma)+ 3\nu \Gamma \gamma > 0$; second, $A= - (\sigma_1+\sigma_2+\sigma_3)>0$; third, $B-A= \sigma_1 \sigma_2+\sigma_1 \sigma_3 + \sigma_3 \sigma_2>0$.

There are also cases related to the special choice of initial data $W(1)$, $E(1)$, as well as, of the choice of parameters $K_0$, $\nu$, $\Gamma$, $\gamma$. For instance, if we deal with the situation indicated as ${\it (i)}$ but now $C_1 =0$ due to specific choice of $W(1)$, $E(1)$, (see (\ref{82})), we obtain the situation  ${\it (ii)}$ or ${\it (iii)}$.

\subsection{Solutions corresponding to the positive discriminant, $\Delta  > 0$}

\subsubsection{The structure of the exact solution}

Now one root, say $\sigma_1$, is real and $\sigma_{2,3}$ are complex conjugated:
\begin{equation}
\sigma_1 = -\frac{A}{3} + ({\cal U} + {\cal V})  \,, \quad
\sigma_{2,3}=  \alpha \pm i \beta \,, \quad \alpha \equiv -\frac{A}{3} -\frac12 ({\cal U} + {\cal V}) \,, \quad
\beta \equiv  \frac{\sqrt3}{2} ({\cal U} - {\cal V}) \,,
\label{91}
\end{equation}
where the auxiliary real parameters ${\cal U}$ and ${\cal V}$
\begin{equation}
{\cal U} \equiv \left[- \frac{q}{2} + \sqrt{\Delta} \right]^{\frac13} \,, \quad {\cal V} \equiv  \left[- \frac{q}{2} - \sqrt{\Delta} \right]^{\frac13}
\label{925}
\end{equation}
are chosen so that ${\cal U}{\cal V}=-\frac{p}{3}$.
Similarly to the case with negative discriminant, we obtain the DE energy density scalar
\begin{equation}
W(x) = C_1 x^{\sigma_1} + x^{\alpha}\left[ C_2 \cos{\beta\log{x}} + C_3 \sin{\beta\log{x}} \right] \,,
\label{92}
\end{equation}
the DM energy density
$$
K_0 E(x)  = C_1 x^{\sigma_1}\left[\sigma^2_1 + \sigma_1 (\nu + 3\Gamma) + (K_0+3 \nu \Gamma)\right]   +
$$
$$
 + x^{\alpha}\left\{
 \left[\alpha^2 -\beta^2 + \alpha (\nu + 3\Gamma) + (K_0+3 \nu \Gamma)\right] \left[C_2 \cos{\beta\log{x}} + C_3 \sin{\beta\log{x}} \right] +
 \right.
 $$
 \begin{equation}
 \left.
 +\beta(2\alpha + \nu + 3 \Gamma) \left[C_3 \cos{\beta\log{x}} - C_2 \sin{\beta\log{x}}\right] \right\} \,,
\label{93}
\end{equation}
where
\begin{equation}
C_1 = \frac{K_0 E(1)+ W(1)\left[(\alpha+3\Gamma)^2+\beta^2 - K_0 \right]}{\left[(\sigma_1-\alpha)^2 + \beta^2\right]} \,,
\label{ccc1}
\end{equation}
\begin{equation}
C_2 = \frac{-K_0 E(1)+ W(1)\left[(\sigma_1-\alpha)^2-(\alpha+3\Gamma)^2+ K_0 \right]}{\left[(\sigma_1-\alpha)^2 + \beta^2\right]} \,,
\label{ccc2}
\end{equation}
\begin{equation}
C_3 = \frac{K_0 (\alpha-\sigma_1)E(1)+ W(1)\left\{(\sigma_1-\alpha)K_0 + (\sigma_1+3\Gamma)\left[(\alpha+3\Gamma)(\alpha-\sigma_1)-\beta^2 \right]\right\}}{\beta \left[(\sigma_1-\alpha)^2 + \beta^2\right]} \,.
\label{ccc2}
\end{equation}
The square of the Hubble function can be extracted from the formula
$$
\frac{3K_0}{\kappa}\left[ H^2(x)-\frac{\Lambda}{3}\right] =  C_1 x^{\sigma_1}\left[2K_0+\sigma^2_1 + \sigma_1 (\nu + 3\Gamma) + 3 \nu \Gamma \right] +
$$
$$
+ x^{\alpha}\left\{
 \left[\alpha^2 -\beta^2 + \alpha (\nu + 3\Gamma) + 2K_0+3 \nu \Gamma\right] \left[C_2 \cos{\beta\log{x}} + C_3 \sin{\beta\log{x}} \right]
 + \right.
 $$
 \begin{equation}
 \left.
 +\beta(2\alpha + \nu + 3 \Gamma) \left[C_3 \cos{\beta\log{x}} - C_2 \sin{\beta\log{x}}\right] \right\}
\,.
\label{941}
\end{equation}

\subsubsection{Admissible asymptotic regimes}

Clearly, all three asymptotic regimes: the Big Rip, Pseudo-Rip, power-law expansion, mentioned above, also  can be realized in this submodel. However, three new elements can be added into the catalog of possible regimes.

\vspace{3mm}
\noindent
{\it (i)}
The first new regime can be indicated as a quasi-periodic expansion; it can be realized when $\sigma_1=0$, $\alpha$ is negative, and $H^2_{\infty}> |h|$. The square of the Hubble function can be now rewritten as follows:
\begin{equation}
H^2 \to H^2_{\infty} + h x^{-|\alpha|}\sin{[\beta \log{x}+ \psi]} \,.
\label{mn1}
\end{equation}
Asymptotically, the Universe expansion tends to the Pseudo Rip regime, however, this process has quasi-periodic features.

\vspace{3mm}
\noindent
{\it (ii)}
The second new regime relates to $\sigma_1=0$, $\alpha=0$ and $H^2_{\infty}> |h|$. The square of the Hubble function, the DE and DM energy densities  become now periodic functions (see, e.g., (\ref{mn1}) with $\alpha=0$).

\vspace{3mm}
\noindent
{\it (iii)}
The third regime is characterized by the following specific feature: $H^2$ takes zero value at finite $x=x_{*}$. This regime can be effectively realized in two cases: first, when $\sigma_1=0$, $\alpha<0$ and $H^2_{\infty}< |h|$;
second, when $\sigma_1=0$, $\alpha>0$. In both cases the size of the Universe is fixed by the specific value of the scale factor $a^{*} = a(t_0) x_{*}$.

\subsection{Solutions corresponding to the vanishing discriminant, $\Delta = 0$}

\subsubsection{Two roots coincide, $q\neq 0$}

It is the case, when all roots are real, but two of them coincide:
\begin{equation}
\sigma_1 = -\frac{A}{3} + 2 \left(- \frac{q}{2} \right)^{\frac13}  \,, \quad
\sigma \equiv \sigma_{2}= \sigma_{3} =  -\frac{A}{3} - \left(- \frac{q}{2} \right)^{\frac13}  \,.
\label{121}
\end{equation}
The DE and DM energy-density scalars contain logarithmic functions
\begin{equation}
W(x) = C_1 x^{\sigma_1} + x^{\sigma}\left[C_2 + C_3 \log{x} \right] \,,
\label{212}
\end{equation}
$$
K_0 E(x)  = C_1 x^{\sigma_1}\left[\sigma^2_1 + \sigma_1 (\nu + 3\Gamma) + (K_0+3 \nu \Gamma)\right]   +
$$
\begin{equation}
+ x^{\sigma}\left\{(C_2+ C_3 \log{x})\left[\sigma^2 + \sigma(\nu + 3\Gamma) + (K_0 + 3\nu \Gamma) \right] + C_3\left(2\sigma + \nu + 3\Gamma\right)\right\}
   \,,
\label{932}
\end{equation}
where the constants of integration are
\begin{equation}
C_1 = \frac{K_0 E(1) + W(1)\left[(\sigma+ 3\Gamma)^2-K_0 \right]}{(\sigma_1-\sigma)^2}
   \,,
\label{ccc21}
\end{equation}
\begin{equation}
C_2 = - \frac{K_0 E(1) + W(1)\left[(\sigma+ 3\Gamma)^2-K_0 - (\sigma_1-\sigma)^2\right]}{(\sigma_1-\sigma)^2}
   \,,
\label{ccc22}
\end{equation}
\begin{equation}
C_3 = \frac{K_0 E(1) + W(1)\left[(\sigma+ 3\Gamma)(\sigma_1+ 3\Gamma)-K_0 \right]}{(\sigma-\sigma_1)}
   \,.
\label{ccc23}
\end{equation}
The square of the Hubble function is presented as follows:
$$
H^2(x) = \frac{\Lambda}{3} {+} \frac{\kappa}{3K_0}\left\{
C_1 x^{\sigma_1}\left[\sigma^2_1 {+} \sigma_1 (\nu {+} 3\Gamma) {+} (2K_0{+}3 \nu \Gamma) \right]
{+}
C_2 x^{\sigma}\left[\sigma^2 {+} \sigma (\nu {+} 3\Gamma) {+} (2K_0{+}3 \nu \Gamma) \right] {+}
\right.
$$
\begin{equation}
\left.
+
C_3 x^{\sigma}\left[\log{x} \left(\sigma^2 + \sigma (\nu + 3\Gamma) + (2K_0+3 \nu \Gamma)\right)+ (2\sigma + \nu + 3\Gamma) \right]
\right\}   \,.
\label{H3}
\end{equation}

\subsubsection{Three roots coincide, $q = 0$}

Now all the roots coincide
\begin{equation}
\sigma_1 = \sigma_{2}= \sigma_{3} =  -\frac{A}{3}= \sigma \,.
\label{221}
\end{equation}
The DE and DM energy-density scalars, the square of the Hubble function contain logarithmic function and its square
\begin{equation}
W(x) =  x^{\sigma}\left[C_1 + C_2 \log{x} + C_3 \log^2{x} \right] \,,
\label{312}
\end{equation}
$$
K_0 E(x) =  x^{\sigma}\left\{ \left(C_1 + C_2 \log{x} + C_3 \log^2{x} \right)\left[\sigma^2+ \sigma(\nu + 3\Gamma) + K_0 + 3\nu \Gamma\right] +
\right.
$$
\begin{equation}
\left.
+ (C_2 +2C_3 \log{x}) (2\sigma+\nu + 3\Gamma) + 2C_3 \right\} \,,
\label{313}
\end{equation}
$$
H^2(x) = \frac{\Lambda}{3} + \frac{\kappa}{3K_0} x^{\sigma} \left\{
\left(C_1+ C_2 \log{x}+ C_3 \log^2{x}\right) \left[\sigma^2 + \sigma (\nu + 3\Gamma) + (2K_0+3 \nu \Gamma) \right]
+
\right.
$$
\begin{equation}
\left.
+
(2\sigma + \nu + 3\Gamma) \left(C_2 + 2C_3 \log{x}\right) + 2C_3
\right\}   \,,
\label{H4}
\end{equation}
\begin{equation}
C_1 = W(1) \,, \quad C_2 = - W(1)(\sigma+3\Gamma) \,, \quad C_3 = \frac12 \left\{K_0 E(1) + W(1)[(\sigma+3\Gamma)^2-K_0] \right\} \,.
\label{413}
\end{equation}

\subsubsection{Admissible asymptotic regimes}

Since the Hubble function contains now the logarithmic terms  $\log{x}$ and $\log^2{x}$, a new asymptotic regime, the so-called Little Rip, is possible. In the case of Little Rip we obtain that asymptotically
$H(t) \to \infty$ and $a(t)\to \infty$, however, the infinite values can be reached during the infinite time interval only.

\section{Three examples of explicit model analysis}

As a preamble, we would like to recall that the set of model parameters ($\Gamma$, $\gamma$, $K_0$, $\nu$, $\Lambda$) is adequate for the procedure of fitting of the acceleration parameter $-q(T) \simeq 0.55$ and of the factor $\frac{E(T)}{W(T)} \simeq \frac{23}{73}$. Nevertheless, we do not perform this procedure in this paper, and do not accompany this procedure by the detailed plots of $q(t)$, $\omega(t)$, $H(t)$, etc. However, we think that for demonstration of analytical capacities of our new model, it is interesting to consider some exact solutions obtained for the set of parameters specifically chosen. Of course, when we introduce the model parameters "by hands", we restricts the time interval, on which the solution is physically motivated and is mathematically adequate. For instance, the super-inflationary solution discussed below can be applicable for the early Universe, but is not appropriate for the late-time period. Nevertheless, the presented exact solutions seem to be intriguing.

\subsection{First explicit submodel, $\Delta<0$, $q=0$ and $\Lambda=0$; how do the initial data correct the Universe destiny? }

For illustration, let us consider the case with the following set of parameters:
\begin{equation}
\Lambda=0 \,, \quad \Gamma=0 \,, \quad \nu = \frac32 \gamma \,, \quad K_0 = - \frac94 \gamma^2  \,.
\label{101}
\end{equation}
Let us recall that for $\Gamma=0$ according to (\ref{31}) we obtain $P=-W$, i.e., the pressure typical for the Dark Energy. One deals with the Cold Dark Matter, when $\gamma=1$; generally, $\gamma \geq 1$. The (\ref{52}) and (\ref{63}) yield
\begin{equation}
q=0 \,, \quad \varphi = \frac{\pi}{2} \,, \quad p=- \frac{27}{4} \gamma^2 \,, \quad B=A = \frac92 \gamma \,,  \quad  D= - \frac{27}{4} \gamma^3  \,, \quad \Delta = - \left(\frac{3\gamma}{2}\right)^6 <0\,.
\label{102}
\end{equation}
Thus, for $\gamma>0$ one root of  the characteristic equation is positive, and other two are negative:
\begin{equation}
 \sigma_1 = \frac32 \gamma (\sqrt3 -1) >0 \,, \quad  \sigma_2 = -\frac32 \gamma (\sqrt3 +1) <0 \,, \quad \sigma_3 = - \frac32 \gamma <0 \,.
\label{103}
\end{equation}
The constants of integration are, respectively,
\begin{equation}
C_1= \frac16 \left[(2{+}\sqrt3) W(1)  {-}  E(1) \right] \,, \quad  C_2= \frac16 \left[(2{-}\sqrt3)W(1)  {-} E(1) \right]\,, \quad C_3= \frac13 \left[W(1)  {+} E(1)\right]
\,.
\label{104}
\end{equation}
Clearly, there are three principal situations, which correspond to three ranges of values of the initial parameter $\omega(1) = \frac{E(1)}{W(1)}$.

\vspace{3mm}
\noindent
{\it (i)}
When $\omega(1)=2+\sqrt3$, i.e., $C_1=0$, and the growing mode is deactivated, the DE energy density, DM energy density
take, respectively, the form
\begin{equation}
W(x)=
\frac{W(1)}{\sqrt3} x^{-\frac32 \gamma} \left[(\sqrt3+1) - x^{-\frac{3\sqrt3}{2} \gamma} \right] \geq 0
\,,
\label{1052}
\end{equation}
\begin{equation}
E(x)= \frac{E(1)}{\sqrt3} x^{-\frac32 \gamma} \left[(\sqrt3-1) + x^{-\frac{3\sqrt3}{2} \gamma} \right] \geq 0
\,.
\label{1051}
\end{equation}
The function $\omega(x)=\frac{E(x)}{W(x)}$, which is  given by
\begin{equation}
\omega(x)= (2+\sqrt3)\frac{\left[(\sqrt3-1) + x^{-\frac{3\sqrt3}{2} \gamma}\right]}{\left[(\sqrt3+1) - x^{-\frac{3\sqrt3}{2} \gamma}\right]} \,,
\label{g1}
\end{equation}
is positive and monotonic; it starts from the value $\omega(1){=} 2{+}\sqrt3$ and tends asymptotically to $\omega(\infty){=}1$. In other words, the energy density of the Dark Matter tends to the  energy density of the Dark Energy due to the interaction of the rheological type.
The square of the Hubble function is also non-negative:
\begin{equation}
H^2(x)=  \frac{\kappa W(1) (\sqrt3+1)}{3\sqrt3} x^{-\frac32 \gamma} \left(2+ x^{-\frac{3\sqrt3}{2} \gamma} \right) \geq 0
\,.
\label{1053}
\end{equation}
The scale factor $a(t)$ can be found from the quadrature:
\begin{equation}
\sqrt{\frac{2\kappa W(1) (\sqrt3+1)}{3\sqrt3}}(t-t_0) = \int_1^{\frac{a(t)}{a(t_0)}} \frac{dx x^{\frac{3\gamma}{4}-1}}
{\sqrt{1+ \frac12 x^{-\frac{3\sqrt3}{2} \gamma}}}
\,.
\label{1054}
\end{equation}
In the asymptotic regime the scale factor behaves as $a(t) \propto t^{\frac{4}{3\gamma}}$, and the Hubble function $H(t)$ tends to zero as $H(t) \simeq \frac{4}{3\gamma t}$.

\vspace{3mm}
\noindent
{\it (ii)}
When $0<\omega(1)<2+\sqrt3$, i.e., $C_1>0$, the integral $\int_1^{\infty} \frac{dx}{xH(x)}$ converges, so the scale factor $a(t)$ reaches infinite value at finite value of the cosmological time. The growing mode, which relates to the positive root $\sigma_1$, become the leading mode, and we obtain the model of the Big Rip type.

\vspace{3mm}
\noindent
{\it (iii)}
When $\omega(1)>2+\sqrt3$, i.e., $C_1<0$, we obtain the model in which the square of the Hubble function takes zero value at some finite time moment. In other words, the Universe expansion stops, the Universe volume becomes finite.

\subsection{Second and third explicit submodels: The case $\Delta=0$ and $q=0$}

For illustration we consider the model, in which all three roots coincide and are equal to zero, $\sigma_1=\sigma_2=\sigma_3=0$. Equivalently, we assume that the characteristic equation takes the form $\sigma^3=0$, and, thus, $A=0$, $B=0$, $D=0$. Only one set of model parameters admits such solution, namely
\begin{equation}
\gamma+\Gamma=0\, \quad  \nu = 0 \,, \quad K_0 = \frac92 \Gamma^2  \,.
\label{exp01}
\end{equation}
In particular, this model covers the case, when $\gamma=1$ and $\Gamma=-1$, i.e., the Dark Matter is pressureless, $\Pi=0$, and the DE pressure is described by the equation of state $P=-2W$.
For this set of guiding parameters we obtain
\begin{equation}
W(x)= W(1)\left(1- 3 \Gamma \log{x} \right) + \frac94 \Gamma^2\left[W(1)+E(1) \right] \log^2{x}\,,
\label{exp011}
\end{equation}
\begin{equation}
E(x)= E(1)\left(1- 3 \gamma \log{x} \right) + \frac94 \gamma^2\left[W(1)+E(1) \right] \log^2{x}\,.
\label{exp0119}
\end{equation}
The square of the Hubble function is presented by the formula
\begin{equation}
H^2(x)= \frac{\Lambda}{3} + \frac{\kappa}{3} \left(1+\frac92 \Gamma^2\log^2{x} \right)\left[W(1)+E(1) \right]  + \kappa \gamma \left[W(1)-E(1) \right] \log{x}  \,,
\label{exp0118}
\end{equation}
and the scale factor can now be found in elementary functions from the integral
\begin{equation}
(t-t_0) = \int^{\log{\frac{a(t)}{a(t_0)}}}_0 \frac{dz}{\sqrt{ \frac{\Lambda}{3} + \frac{\kappa}{3} \left(1+\frac92 \Gamma^2 z^2 \right)\left[W(1)+E(1) \right] + \kappa \gamma \left[W(1)-E(1) \right] z}} \,.
\label{exp0117}
\end{equation}
However, the integration procedure is faced with two principally different cases, $W(1)+E(1) = 0$, and $W(1)+E(1) \neq 0$. Let us consider them separately.

\subsubsection{The case $W(1)+E(1) = 0$: Solution of the Bounce type}

While this case seems to be exotic (one of the energy densities should be negative), it is interesting to study this case in detail. First, we fix that $\gamma>0$, $W(1)>0$. Then, integration gives immediately
\begin{equation}
a(t) = a(t_{*}) \exp{\left[\frac12 \gamma \kappa W(1) (t-t_{*})^2 \right]} \,,
\label{exp13}
\end{equation}
where the following auxiliary parameters are introduced
\begin{equation}
a(t_{*}) = a(t_0)\exp{\left[- \frac{\Lambda}{6\kappa W(1)\gamma} \right]}\,, \quad
t_{*} = t_0 -  \frac{\sqrt{\frac{\Lambda}{3}}}{\kappa W(1)\gamma} \,.
\label{2exp138}
\end{equation}
In terms of cosmological time the Hubble function is the linear one:
\begin{equation}
H(t) = \kappa W(1)\gamma(t-t_{*}) \,.
\label{exp14}
\end{equation}
The corresponding acceleration parameter
\begin{equation}
-q(t)= 1 + \frac{1}{\kappa W(1)\gamma (t-t_{*})^2}
\label{exp147}
\end{equation}
tends to one asymptotically at $t \to \infty$. In the work \cite{Arc1} the solution of this type was indicated as anti-Gaussian solution; also this solution is known as bounce (see, e.g., \cite{bounce}).

The DE and DM energy densities behave as quadratic functions of cosmological time:
\begin{equation}
\frac{W(t)}{W(1)}= \frac32 \gamma^2 \kappa W(1)(t-t_{*})^2 + 1 - \frac{\Lambda}{2\kappa W(1)}\,,
\label{bounceW}
\end{equation}
\begin{equation}
\frac{E(t)}{W(1)}= \frac32 \gamma^2 \kappa W(1)(t-t_{*})^2 - 1 - \frac{\Lambda}{2\kappa W(1)}\,.
\label{bounceE}
\end{equation}
It is interesting to mention that the rates of evolution of the DE and DM energy density scalars coincide:
\begin{equation}
\dot{E}(t)= \dot{W}(t) = 3 \gamma^2 \kappa W^2(1)(t-t_{*}) \,.
\label{bounce7}
\end{equation}
Clearly, both functions: $\omega(t)=\frac{E(t)}{W(t)}$ and $-q(t)$ tend asymptotically to one, $\omega(\infty)=1$, $-q(\infty)=1$. The acceleration parameter $-q(t)$ described by the simple monotonic function (\ref{exp147}).

\subsubsection{The case $W(1)+E(1) \neq 0$: Super-inflationary solution}

For illustration we consider the simple submodel with $\Lambda=0$, and assume that at $t=t_0$ the DE and DM energy densities coincide, i.e., $E(1)=W(1)$.
The integration in (\ref{exp0117}) yields now
\begin{equation}
\log{\frac{a(t)}{a(t_0)}} = \frac{\sqrt2}{3\gamma} \sinh{\left[\gamma(t-t_0)\sqrt{3\kappa W(1)}\right]} \,.
\label{super1}
\end{equation}
This solution is of the super-inflationary type; at $t \to \infty$ it behaves as
\begin{equation}
\frac{a(t)}{a(t_0)} = e^{\frac{1}{3\sqrt2 \gamma} e^{\sqrt{3 \kappa W(1)} \ \gamma \ t }} \,.
\label{exp4}
\end{equation}
It can be indicated as a Little Rip according to the classification given in \cite{O1}. Also this solution appears in the model of Archimedean-type interaction between DE and DM \cite{Arc1}.

The DE and DM energy densities behave as follows:
\begin{equation}
\frac{W(t)}{W(1)} = \cosh^2{\left[\gamma(t-t_0)\sqrt{3\kappa W(1)}\right]} + \sqrt2 \sinh{\left[\gamma(t-t_0)\sqrt{3\kappa W(1)}\right]} \,,
\label{super3}
\end{equation}
\begin{equation}
\frac{E(t)}{W(1)} = \cosh^2{\left[\gamma(t-t_0)\sqrt{3\kappa W(1)}\right]} - \sqrt2 \sinh{\left[\gamma(t-t_0)\sqrt{3\kappa W(1)}\right]} \,,
\label{super31}
\end{equation}
so, the function $\omega(t)=\frac{E(t)}{W(t)}$ tends asymptotically to one.
The Hubble function and acceleration parameter are, respectively
\begin{equation}
H(t)= \sqrt{\frac23 \kappa W(1)} \cosh{\left[\gamma(t-t_0)\sqrt{3\kappa W(1)}\right]}\,,
\label{super7}
\end{equation}
\begin{equation}
 -q(t) = 1+ \left(\frac{3\gamma}{\sqrt2} \right)  \frac{\sinh{\left[\gamma(t-t_0)\sqrt{3\kappa W(1)}\right]}}{\cosh^2{\left[\gamma(t-t_0)\sqrt{3\kappa W(1)}\right]}}\,.
\label{exp32}
\end{equation}
When we study the time interval $t \geq t_0$, we see that the function $-q(t)$ starts with $-q(t_{0})=1$, reaches the maximum $-q_{(\rm max)} = 1 + \frac{3\gamma}{2\sqrt2}$ and then tends to one asymptotically, $-q(\infty)=1$.

\section{Discussion}

We established  the model of DE/DM interaction based on the interaction kernel of the Volterra type, as well as, classified and studied the obtained exact solutions. From our point of view, the results are inspiring. Let us explain our optimism.

 1. The model of kernel of the DE/DM interaction, which possesses two extra parameters, $K_0$ and $\nu$, is able to describe many known interesting cosmic scenaria: Big Rip, Little Rip, Pseudo Rip, de Sitter-type expansion; the late-time accelerated expansion of the Universe is the typical feature of the presented model.

 2. When $2K_0 {+} 3 \nu \Gamma \neq 0$, the solution of a new type appears, which is associated with the so-called {\em effective cosmological constant}. Indeed, if the standard cosmological constant vanishes, $\Lambda=0$, we obtain according to (\ref{006}) that the parameter $H_{\infty} \neq 0$ plays the role of an effective Hubble constant. It appears as the result of integration over the whole time interval; it can be associated with the memory effect produced by the DE/DM interaction; we can introduce the effective cosmological constant $\Lambda_{*}\equiv 3H^2_{\infty}$, which appears just due to the interaction in the Dark sector of the Universe.

3. The regular bounce-type (see (\ref{exp13})) and super-inflationary (see  (\ref{super1})) solutions appear, when the characteristic polynomial of the key equation admits three coinciding roots $\sigma=0$. Both exact solutions belong to the class of solutions describing the Little Rip scenaria.

4. The model of the DE/DM coupling based on the Volterra-type interaction kernel can solve the Coincidence problem. Indeed, the asymptotic value $\omega(\infty)$ of the function $\omega(x) = \frac{E(x)}{W(x)}$ is predetermined by the choice of parameters $K_0$ and $\nu$ entering the integral kernel (\ref{21}), (\ref{22}). Even if the initial value $E(1)$ of the Dark Matter energy density  is vanishing, the final value $E(\infty)$ is of the order of the final value $W(\infty)$ due to the integral procedure of energy redistribution, which is described by the Volterra operator (see, e.g., the example (\ref{q125})). In other words, the DE component of the Dark Fluid transmits the energy to the DM components during the whole evolution time interval, and this action "is remembering" by the Dark Fluid.

5. Optimization of the model parameters $K_0$, $\nu$, $\Gamma$, $\gamma$
using the observational data is the goal of our next work. However, some qualitative comments concerning the ways to distinguish the models of DE/DM interactions can be done based on the presented work. For instance, when one deals with the standard $\Lambda$CDM model, the profile of the energy density associated with the Dark Energy is considered to be the horizontal straight line; the DM energy density profile decreases monotonically, thus providing the existence of some cross-point at some finite time moment. For this model the time derivative $\dot{W}(t)$ vanishes, so that $\dot{W}(t)=0$ and $\dot{E}(t) \neq 0$ never coincide. In the model under discussion, the profiles $E(t)$ and $W(t)$ do not cross; these quantities tend to one another asymptotically. As for the rates of evolution, the quantities $\dot{W}(t)$ and $\dot{E}(t)$ can coincide identically (see, e.g., (\ref{bounce7})), or can tend to one another asymptotically. In other words, one can distinguish the models of DE/DM interaction if to analyze and compare the rates of evolution of the DE and DM energy density scalars.

\acknowledgments{The work was supported by Russian Science Foundation (Project No. 16-12-10401), and, partially, by the Program of Competitive Growth
of Kazan Federal University.}


\begin{thebibliography}{99}


\bibitem{DM1}Turner, M.S. The dark side of the universe: from Zwicky to accelerated expansion. {\em Phys. Rept.} {\bf 2000}, {\em 333, 334}, 619-635.

\bibitem{DE4} Peebles, P.J.E.; Ratra B. The Cosmological Constant and Dark Energy. {\em Rev. Mod. Phys.} {\bf 2003}, {\em 75}, 559-606.

\bibitem{DMDE1} Sahni, V. Dark Matter and Dark Energy. {\em  Lect. Notes Phys.} {\bf 2004}, {\em 653}, 141-180.

\bibitem{DE5}  Copeland, E.J.;  Sami, M.; Tsujikawa, S. Dynamics of dark energy. {\em Int. J. Mod. Phys. D} {\bf 2006}, {\em 15}, 1753-1935.

\bibitem{DE6} Sahni, V.;   Starobinsky, A. Reconstructing Dark Energy. {\em Int. J. Mod. Phys. D} {\bf 2006}, {\em 15}, 2105-2132.

\bibitem{O4} Capozziello, S.; Nojiri, S.; Odintsov, S.D. Unified phantom cosmology: inflation, dark energy and dark matter under the same standard.
{\em Phys. Lett. B } {\bf 2006}, {\em 632}, 597-604.

\bibitem{O3} Nojiri, S.; Odintsov S.D. Introduction to Modified Gravity and Gravitational Alternative for Dark Energy.
{\em Int. J. Geom. Meth. Mod. Phys.} {\bf 2007}, {\em 4}, 115-146.

\bibitem{DE7}  Frieman,  J.; Turner, M.; Huterer, D. Dark Energy and the Accelerating Universe. {\em Ann. Rev.
Astron. Astrophys.} {\bf 2008}, {\em 46}, 385-432.

\bibitem{DE8} Padmanabhan, T. Dark Energy and Gravity. {\em Gen. Rel. Grav.} {\bf 2008}, {\em 40}, 529-564.

\bibitem{NM8} Bamba, K.;  Odintsov, S.D.   Inflation and late-time cosmic acceleration in non-minimal Maxwell-$F(R)$
gravity and the generation of large-scale magnetic fields. {\em JCAP} {\bf 2008}, {\em 0804}, 024-1-024-21.

\bibitem{O1} Nojiri, S.; Odintsov, S.D. Unified cosmic history in modified gravity: from F(R) theory to Lorentz non-invariant models. {\em Phys. Rept.} {\bf 2011}, {\em 505}, 59-144.

\bibitem{bounce} Nojiri, S.; Odintsov, S.D.; Oikonomou, V.K. Modified Gravity Theories on a Nutshell: Inflation, Bounce and Late-time Evolution. {\em Phys. Rept.} {\bf 2017}, {\em 692}, 1-104.

\bibitem{O2} Bamba, K.; Capozziello, S.; Odintsov, S.D. Dark energy cosmology: the equivalent description via different theoretical models and cosmography tests. {\em Astrophysics and Space Science} {\bf 2012}, {\em 342}, 155-228.

\bibitem{DM2} Del Popolo, A. Non-baryonic dark matter in cosmology. {\em Int. J. Mod. Phys. D} {\bf 2014}, {\em 23}, 1430005-1-1430005-109.

\bibitem{DM3} Yepes, G.; Gottlober, S.; Hoffman Y. Dark matter in the Local Universe. {\em New Astronomy Reviews} {\bf 2014}, {\em 58}, 1-18.

\bibitem{DM4} Zurek, K.M. Asymmetric Dark Matter: Theories, signatures, and constraints. {\em Phys. Rept.} {\bf 2014}, {\em 537}, 91-121.

\bibitem{DMDE4}Gleyzes, J., Langlois, D., Vernizzi, F. A unifying description of dark energy. {\em Int. J. Mod. Phys. D} {\bf 2015}, {\em 23}, 1443010.


\bibitem{CO1} Chimento, L.P.; Jacubi A.S.; Pavon, D.; Zimdahl, W. Interacting quinessence solution to the coincidence problem. {\em Phys. Rev. D} {\bf 2003}, {\em 67}, 083513.

\bibitem{CO2} Scherer, R.J. Phantom Dark Energy, Cosmic Doomsday, and the Coincidence Problem. {\em Phys. Rev. D} {\bf 2005}, {\em 71}, 063519.

\bibitem{CO3} Velten, H.E.S.; vom Marttens, R.F.; Zimdahl, W. Aspects of the cosmological "coincidence problem". {\em Eur. Phys. J.C.} {\bf 2014}, {\em 74}, 3160.

\bibitem{Pavon2}
Wang, B.; Abdalla, E.; Atrio-Barandela, F.; Pavon, D. Dark Matter and Dark Energy Interactions: Theoretical Challenges, Cosmological Implications and Observational Signatures. arXiv: 1603.08299.

\bibitem{i1} Farrar, G.R.; Peebles, P.J.E. Interacting dark matter and dark energy. {\em  Astrophys. J.} {\bf 2004}, {\em 604}, 1-11.

\bibitem{Z} Zimdahl, W. Interacting dark energy and cosmological equations of state. {\em Int. J. Mod. Phys. D}, {\bf 2005}, {\em 14}, 2319-2326.

\bibitem{Pavon1}
Del Campo, S.; Herrera, R.; Pavon, D. Interaction in the Dark Sector. {\em Phys. Rev. D} {\bf 2015}, {\em 91}, 123539.

\bibitem{Arc1} Balakin, A.B.; Bochkarev, V.V. Archimedean-type force in a cosmic dark fluid. I. Exact solutions for the late-time accelerated expansion. {\em Phys. Rev. D} {\bf 2011},
 {\em 83}, 024035-1-024035-12.

\bibitem{Arc2} Balakin, A.B.; Bochkarev, V.V. Archimedean-type force in a cosmic dark fluid. II. Qualitative and numerical study of a multistage universe expansion. {\em Phys. Rev. D} {\bf 2011},
 {\em 83}, 024036-1-024036-15.

\bibitem{Arc3} Balakin, A.B.;  Bochkarev, V.V.  Archimedean-type force in a cosmic dark fluid. III. Big Rip, Little Rip and Cyclic solutions. {\em Phys. Rev. D} {\bf 2013}, {\em 87},
024006-1-024006-16.

\bibitem{Arc4} Balakin, A.B.; Bochkarev, V.V.; Lemos, J.P.S. Light propagation with non-minimal couplings in a two-component cosmic dark fluid with an Archimedean-type force, and unlighted cosmological epochs. {\em Phys. Rev. D} {\bf 2012}, {\em 85}, 064015-1-064015-17.

\bibitem{BD2014}  Balakin, A.B.;  Dolbilova, N.N. Electrodynamic phenomena induced by a dark fluid: Analogs of pyromagnetic, piezoelectric, and striction effects. {\em Phys. Rev. D} {\bf 2014},
{\em 89}, 104012-1-104012-14.

\bibitem{Obzor} Balakin, A.B. Electrodynamics of a CosmicDark Fluid. {\em Symmetry} {\bf 2016}, {\em 8}, 56-1-56-48.

\bibitem{QQ} Jim\'enez, J.B.; Rubiera-Garcia, D.; S\'aez-G\'omez, D.; Salzana V. Cosmological future singularities in interacting dark energy models. {\em Phys. Rev. D} {\bf 2016}, {\em 94}, 123520.

\bibitem{Volterra} Brunner, H. {\em Volterra Integral Equations}; Cambridge University Press: Cambridge, UK, {\bf 2017}.


\end{thebibliography}
\end{document}